\documentclass[aps,prb,reprint,superscriptaddress,amsmath,amssymb]{revtex4-1}

\usepackage{graphicx}
\usepackage{dcolumn}

\usepackage{color}

\begin{document}
\title{ Role of Hydrogen in the Electronic Properties of CaFeAsH-based Superconductors }
\author{Y. N. Huang}
\affiliation{Key Laboratory of Materials Physics,
Institute of Solid State Physics, Chinese Academy of Sciences,
P. O. Box 1129, Hefei 230031, China}
\affiliation{University of Science and Technology of China,
             Hefei, Anhui 230026, China}
\affiliation{Department of Physics, University of California Davis,
	Davis, California 95616, USA}
\author{D. Y. Liu}
\affiliation{Key Laboratory of Materials Physics,
Institute of Solid State Physics, Chinese Academy of Sciences,
P. O. Box 1129, Hefei 230031, China}
\author{L. J. Zou}
\email{zou@theory.issp.ac.cn}
\affiliation{Key Laboratory of Materials Physics,
             Institute of Solid State Physics, Chinese Academy of Sciences,
             P. O. Box 1129, Hefei 230031, China}
\affiliation{University of Science and Technology of China,
             Hefei, Anhui 230026, China}
\author{W. E. Pickett}
\email{wepickett@ucdavis.edu}
\affiliation{Department of Physics, University of California Davis,
    Davis, California 95616, USA}

\date{\today}

\begin{abstract}
The electronic and magnetic properties of the new hydride superconductor CaFeAsH,
which superconducts up to 47 K when electron-doped with La,
and the isovalent alloy system CaFeAsH$_{1-x}$F$_x$
are investigated using density functional based methods. 
The $\vec Q = (\pi,\pi,0)$ peak of the nesting function $\xi(\vec q)$ is found 
to be extremely strong and sharp, and additional structure in 
$\xi(\vec q)$ associated with the near-circular Fermi surfaces (FSs) that may impact low
energy excitations is quantified. The unusual band introduced by H,
which shows strong dispersion  perpendicular
to the FeAs layers, is shown to be connected to a peculiar van Hove singularity
just below the Fermi level. This band provides a three dimensional electron ellipsoid
Fermi surface not present in other Fe-based superconducting materials nor in CaFeAsF. 
Electron doping by 25\% La or Co has minor affect on this ellipsoid 
Fermi surface, but suppresses FS nesting strongly, consistent with the viewpoint
that eliminating strong nesting and the associated magnetic order allows high 
T$_c$ superconductivity to emerge. Various aspects of the isovalent alloy
system CaFeAsH$_{1-x}$F$_x$ and means of electron doping are discussed in terms
of influence of incipient bands.

\end{abstract}


\maketitle

\section{Introduction}

Identifying and studying similar materials that display very different behavior
has been an active approach to understanding the origin of high temperature
(T$_c$) superconducting Fe-based pnictides and chalcogenides, and is expected
to promote the quest for new and possibly higher T$_{c}$ superconductors.
In this class of materials, elemental substitution can be done in an isovalent
manner to provide (seemingly) minor changes, while aliovalent substitution
supplies either electron or hole carriers to the Fe $3d$ 
bands.\cite{Rotter2008, Fu2008, Wu2008, Zhu2009, Chen2009}
Each method can provide insight not otherwise obtainable.

While there are now several classes of Fe-based superconductors, and most are well studied,
the recently synthesized CaFeAsH\cite{Muraba2014, Hosono2014, Hanna2013, Hanna2011}, 
with its H$^{-}$ anion\cite{Takahashi2012} in the
blocking layer, adds a new dimension to the phenomenon.
Hosono {\it et al.}\cite{Takahashi2012}
showed that hydrogen in this material can form a solid solution in
any ratio with F$^{-}$, making CaFeAsF$_{1-x}$H$_{x}$ an isovalent, isostructural
system in which specific differences can be varied continuously.
Though Muraba {\it et al.}\cite{Hosono2014} have emphasized some
important differences from the oxide analogs
in the electronic structure, the differences arising from H versus F
have not been studied thoroughly.
Such isovalent sister compounds provide means to evaluate the effects
of specific, seemingly small differences, such has been exploited 
in the LiFeAs, NaFeAs, and MgFeGe
compounds, which likewise are isoelectronic and isostructural with extremely similar band
structures,\cite{Ferber2012,Clarke2009,Rhee2013,Jeschke2013}
but differing properties. In fact, MgFeGe is not superconducting at all.
Though non-superconducting at ambient pressure, both CaFeAsH and CaFeAsF display
pressure-driven superconductivity with T$_c$ up to 25-28 K at the superconducting
onset at 3-5 GPa, beyond which T$_c$ decreases with additional pressure.\cite{Okuma2012} 

Impressively, indirect
electron-doping by La on the Ca site of CaFeAsH  produces
superconductivity\cite{Muraba2014} with T$_{c}^{max}$=47 K, in the regime of the
rare earth 1111 compounds with the highest T$_c$'s (up to 56 K). 
This high T$_c$ can be contrasted with ``direct'' electron doping 
by Co-substitution of Fe, which attains\cite{Hosono2014} only T$_{c}^{max}$=23 K.
Also noteworthy is that in both cases there is no
superconducting dome -- the  ``onset T$_c$'' (some samples are evidently not
homogeneous) is almost constant\cite{Muraba2014,Cheng2014} from the
appearance of superconductivity at a doping level at $x$=0.07-0.08 to
the maximum doping studied, $x$=0.3-0.4. 

There are strong parallels to
another two dimensional (2D) electron-doped system $A_x$$M$NCl, 
where $M$ = Zr (respectively Hf), $A$ is
an alkali metal, with a T$_c$ of  15 K (respectively 26 K). In these two
compounds T$_c$ is independent of doping level\cite{Taguchi2006} except at the low end,
at the onset of superconduction where T$_c$ is somewhat higher. This indifference of T$_c$
to the evolving electronic structure has been discussed,\cite{Botana2014}
and this same indifference in CaFeAsH raises important questions, since
there is a great deal of evidence in other Fe-based superconductors
that properties including T$_c$ are sensitive to small changes in electronic structure. 
There are differences in T$_{c}$ and other properties with
chemical variation on the H/F site, in addition to the doping 
differences, that so far
have no explanation.

Incorporation of H as an anion in the blocking layer and the realization 
of the superconductivity as high as 47 K invites certain questions. First, within
the same 1111 phase, why does T$_{c}$ increase so much with La doping after (LaO)$^{+}$
layers are replaced by isovalent (CaH)$^{+}$ layers? 
Second, in comparison with CaFeAsF, what is the role of H
in stabilizing such a high T$_{c}$? A third question is why the maximal 
superconducting critical temperatures in La-doped and Co-doped CaFeAsH 
samples (both electron doped) are so different. These are the sort of
issues we try to shed light on.

Using methods similar to those we use (see below), Muraba {\it et al.}\cite{Hosono2014} 
and Wang and collaborators\cite{Wang2015}
have identified the main differences in the band structures of CaFeAsH and
CaFeAsF: the former has an additional Fermi surface arising from a band
that is displaced from its position in the F compound and in other 1111 compounds.
While it is in all cases a primarily Fe $3d$ band, the H $1s$ states are less
strongly bound than the F $2p$ states and evidently provide a propitious pathway for Fe-As-H-As-Fe
hopping from FeAs layer to FeAs layer. Fermi surface nesting was discussed
but the difference between the H and F compounds was not quantified.

These differences, and the electronic structure to be analyzed here, 
call to mind the recent discussion on the possible
influence of incipient bands in Fe-based superconductors. Incipient bands
are those that do not actually cross the Fermi energy (E$_F$) and thus do not
have an associated Fermi surface, but are near enough to the Fermi level
(within a pairing cutoff energy) to contribute to pairing and superconductivity.
Chen {\it et al.},\cite{Chen2015} who provide an excellent overview of the
discussion, conclude that incipient bands may indeed play a role in
pairing in LiFeAs and in FeSe monolayers on SrTiO$_3$ substrates. 
The difference in the electronic structures at or near the Fermi level,
and the associated (magnetic) fluctuations, in
these H and F compounds must account for their differing properties,
and the incipient band viewpoint may be a useful one to consider. 

In this paper first principles methods, described in Sec. II, 
are applied to study the electronic structures
and magnetic ground states of CaFeAsH and CaFeAsF, and the non-magnetic
and superconducting electron-doped phases
Ca$_{0.75}$La$_{0.25}$FeAsH and CaFe$_{0.75}$Co$_{0.25}$AsH.
Sec. III contains the main results. It is confirmed that 
CaFeAsH has, in addition to the nearly 
circular, 2D electron and hole cylinders expected of an Fe 1111 compound, 
a strongly $k_z$-dispersive band crossing the Fermi level
(E$_F$) that  leads to a three dimensional (3D) ellipsoidal  Fermi pocket, whose implications
we explore. A specific feature that we quantify is extremely sharp $\vec Q = (\pi,\pi,0)$
nesting (as usual, in units of in-plane lattice constant $a$=1). Several other
differences between the H- and F-compounds are analyzed, including nesting away from the 
nesting peak at $\vec Q$ that may impact low energy properties. 
A short summary is provided
in Sec. IV.

\section{Structure and Methods}
\subsection{Crystal Structure}
The crystal symmetry across the CaFeAsF$_{1-x}$H$_x$ system
is tetragonal $P4/nmm$, comprised of
alternating Ca$_2$H$_2$ and Fe$_2$As$_2$ layers as in Fe-based 1111 compounds.\cite{Kamihara2008}
The hydride and fluoride ions have closed shell, negatively charged ionic
configurations with not greatly differing anionic radius, allowing the 
successful synthesis of all solid solutions.\cite{Hanna2011} As mentioned,
CaFeAsH displays differences compared to CaFeAsF,\cite{Okada2010,Mishra2011} which is part
of the motivation for this study.

The experimental lattice parameters are used in the calculations.
In the $P4/nmm$ phase above the magnetic ordering temperature, 
for CaFeAsH\cite{Hosono2014} $a_t$=3.879 \AA, $c_t$=8.26 \AA.
For CaFeAsF\cite{Matsuishi2009} the lattice parameter $a$=3.878 \AA~is nearly
identical, while $c$=8.593 \AA. Thus the larger F ion, confirmed by charge
density plots (not shown), enlarges the structure only along the $c$ axis.
For CaFeAsH in the magnetically ordered $Cmma$ phase, 
$a$=5.457 \AA = 1.407$a_t$, $b$=5.492 \AA = 1.416$a_t$, and $c$=8.21\AA = 0.994$c_t$.
For Ca$_{0.75}$La$_{0.25}$FeAsH and CaFe$_{0.75}$Co$_{0.25}$AsH in the tetragonal 
phase, the unit cell size is $2a\times b\times c$, $a$=3.883 \AA= 1.001$a_t$, 
$c$=8.20 \AA = 0.993$c_t$.\cite{Cheng2014}

It will be useful to be aware of the distances between the As plane and the
anion plane. For CaFeAsH, $d$(As-H) = 2.71 \AA; for CaFeAsF, $d$(As-F) = 2.89 \AA.
For comparison, in LaFeAsO, the separation is $d$(As-O) = 3.05 \AA. These
differences have the potential to affect $k_z$ dispersion, that is,
the three dimensional character, which is an issue that we study.

\begin{figure}
  \centering
 \includegraphics[width=0.98\columnwidth]{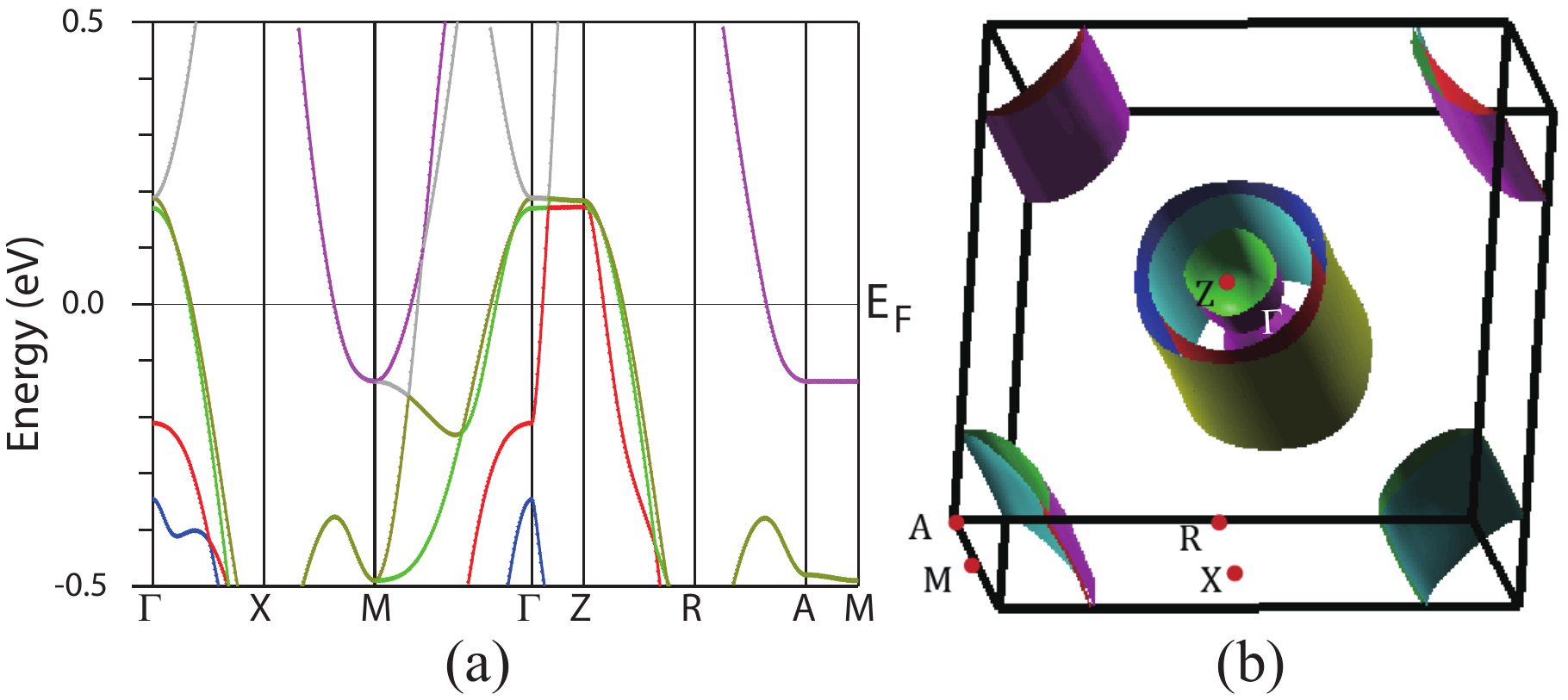}\\
  \caption{(Color online) The (a) electronic band structure and (b) Fermi surfaceis of CaFeAsH in its nonmagnetic phase. In (b) the $\Gamma$ point in the
center of the figure.}\label{Fig1}
\end{figure}

\subsection{Calculational Methods}
The full-potential linearized augmented plane wave
Wien2K package\cite{Blaha2001} has been used for the electronic structure calculations. We use
the Perdew, Burke, and Ernzerhof\cite{Perdew1996} version of the generalized
gradient approximation (GGA) to the exchange-correlation functional within density functional theory.
The sphere radii for Ca, H, Fe, As, F, La and Co are taken as
2.50, 1.40, 2.39, 2.12, 2.20, 2.50 and 2.38 Bohr, respectively. The basis set cut-off parameter
R$_{mt}$$\cdot$K$_{max}$ = 7.0 was found to be sufficient. The number of $\vec k$ points
was typically 3000 for the tetragonal unit cell.

The magnetic order commonly observed in these Fe-based two dimensional
(2D) materials leads us to study
the relative energetics of ordered phases. The primitive tetragonal cell contains two
Fe sites, due to the tetrahedral coordination with As ions. The Fe sublattice is square,
oriented at 45$^{\circ}$ with respect to the conventional cell. One possible magnetic order is
the N\'eel type, in which each Fe on its square sublattice is antialigned with its
four neighbors; this is $\vec q$=0 AFM order (NAFM). 
Stripe AFM (SAFM) order corresponds
to each Fe atom being antiparallel to its neighbors along the $\vec a$ and $\vec b$ axes,
which are second Fe neighbors. SAFM corresponds to $\vec Q = (\pi,\pi,0)$ order, and typically
destroys much of the Fermi surface.
The SAFM unit
cell size is doubled to $\sqrt{2}\times \sqrt{2}\times 1$.

\section{ANALYSIS AND DISCUSSION}
\subsection{Nonmagnetic phase: a new band}

Figure \ref{Fig1} displays the band structure and Fermi surfaces of 
CaFeAsH in the nonmagnetic phase. There are two cylindrical hole bands around the 
$\Gamma$=(0,0,0) point and two cylindrical electron bands crossing E$_F$ near the 
M=($\pi$,$\pi$,0) point, consistent with those presented by Muraba {\it et al.}\cite{Muraba2014}
This band structure is similar to that of the LaFeAsO 1111-type compounds,
except for an additional central Fermi surface around the $Z$ point of the zone. 
The apparent nesting, which we quantify below,
implies that CaFeAsH has a tendency toward the SAFM order phase observed and
obtained computationally in other 1111
compounds. 

The distinctive feature of these bands is the existence of one band along the 
$\Gamma-Z$ direction that shows not only $k_z$ dispersion but a surprisingly large
velocity along $k_z$. This band gives rise to an additional,
ellipsoidal in shape,  hole Fermi surface surrounding the $Z$ point of the zone.
Bands near E$_F$, including the unusual one, have strongly Fe  $d$$_{xy}$ and 
$d$$_{xz},d$$_{yz}$ character, with small As $4p$ character.
This new Fermi surface has similarities to one in a magnetically ordered
phase of LaFePO that is different from the corresponding Fermi surface of 
LaFeAsO in the same magnetic structure.\cite{Seb} The $k_z$ dispersion is however
much less dramatic in LaFePO than in CaFeAsH. 

\subsection{Electronic structure: H versus F}
\begin{figure}
	\centering
	\includegraphics[width=0.98\columnwidth]{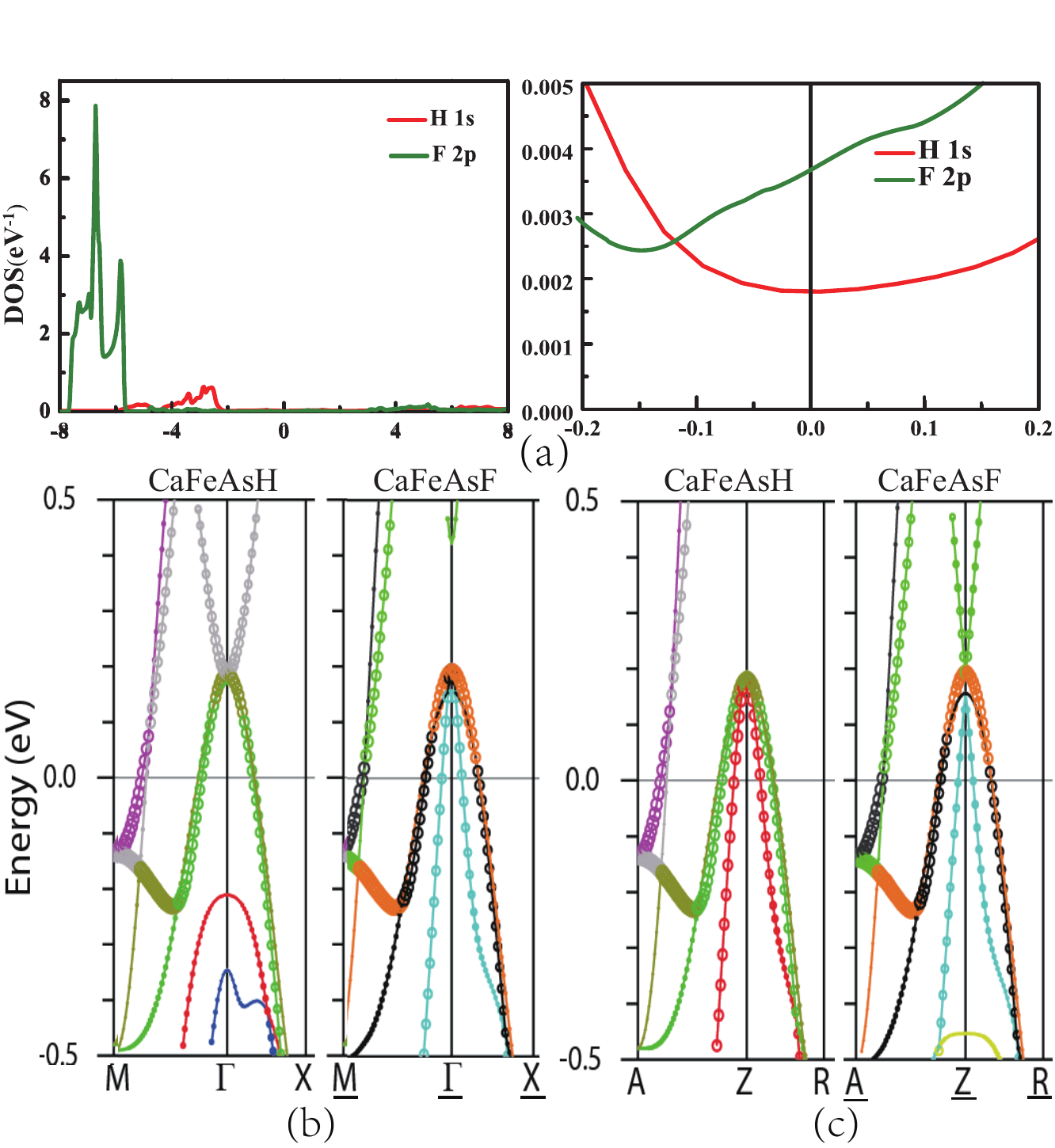}\\
	\caption{(Color online) (a) Left panel: he PDOS of H $1s$ orbital in CaFeAsH and F $2p$ orbitals in CaFeAsF, revealing the substantially larger binding energy of the F states. (a) Right panel: an enlargment of the PDOS around the Fermi energy. (b) and (c) present the 
band structures of both compounds, along the paths (b) M-$\Gamma$-X and (c) A-Z-R at the top of the zone. Paths with underlined letters is for CaFeAsF. The thickness of the bands shows the relative amount of Fe d$_{xz}$+d$_{yz}$ character.}
	\label{Fig2}
\end{figure}
%


Muraba {\it et al.}\cite{Hanna2013} have shown that the F $2p$ and H $1s$ bands lie well below
the Fermi level and are filled, so each is a negatively charged (-1) ion. The
densities of states (DOS) in Fig. \ref{Fig2}(a)
demonstrate that the F states are 2 eV more strongly bound than the H $1s$ states. 
Figure~\ref{Fig2}(a), right panel,  also shows
the corresponding projected density of states (PDOS) near E$_F$. These curves likely reflect not
the atomic states but the tails of neighboring atomic orbitals that extend into the H and
F atomic regions. The point is that these are states near E$_F$ and their distributions
and hybridization are significantly different.
It is this and related differences that we pursue in this subsection.

The Fermi surface differences noted in Sec. III.A do not convey the extent of the differences
in electronic structure caused by H versus F.
~~Figures~\ref{Fig2}(b) and \ref{Fig2}(c) show the band dispersions, weighted by
Fe $d_{xz}$+$d_{yz}$ character, in the M-$\Gamma$-X and the zone top A-Z-R paths.
In CaFeAsF, there are three hole-like bands centered around the
$\Gamma$-Z line, and the bands display no $k_z$ dispersion. 
For CaFeAsH, however, two hole-like bands cross E$_{F}$ along the
M-$\Gamma$-X path while three hole-like bands crosses the E$_{F}$ along the A-Z-R path.
The difference is that in the H compound, a band at $\Gamma$ 0.2 eV below E$_F$ 
in CaFeAsF lies 0.2 eV
above E$_F$ at Z. 

\begin{figure}
	\centering
	\includegraphics[width=0.95\columnwidth]{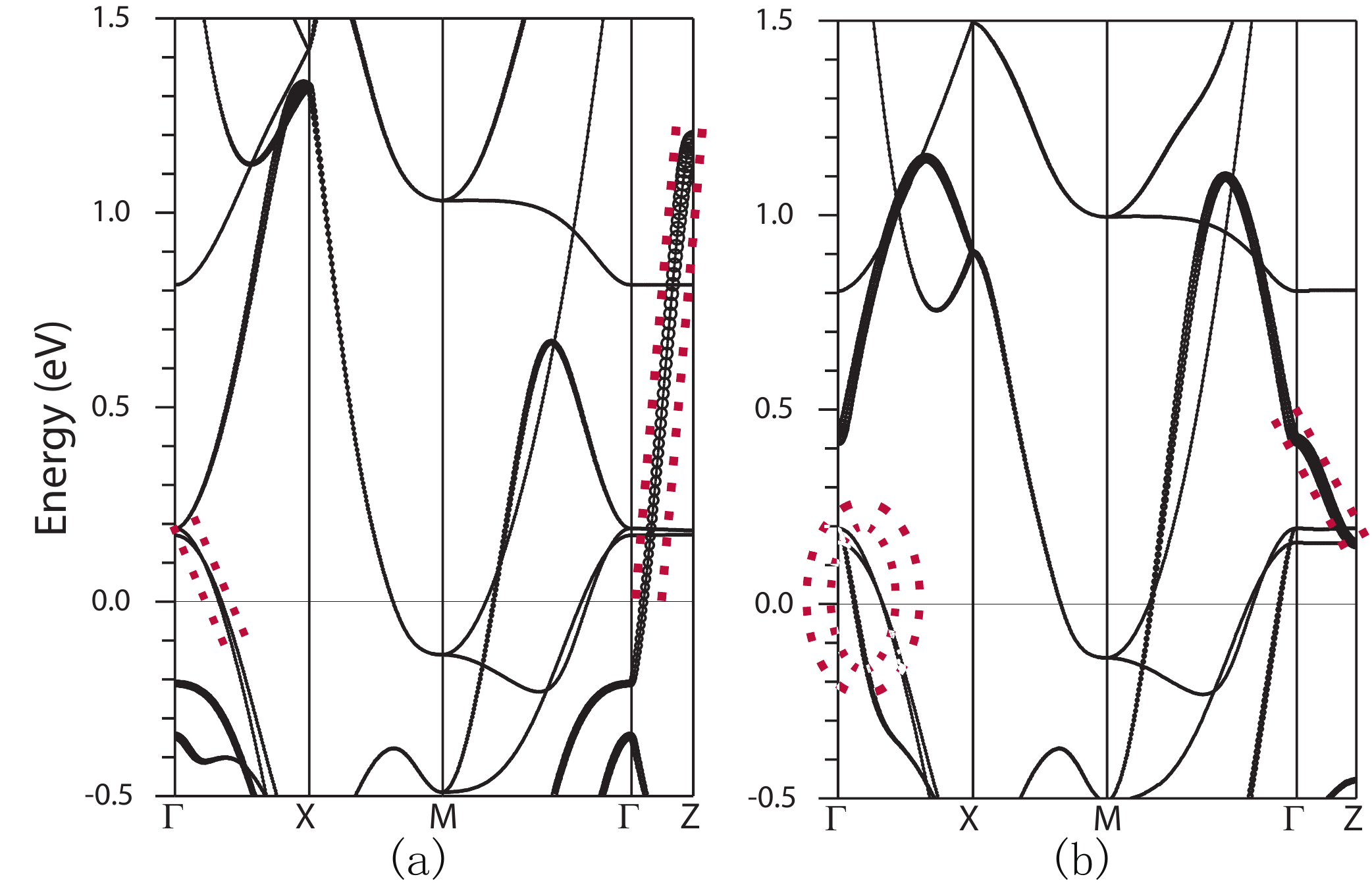}\\
	\caption{(Color online) The contribution of the arsenic atomic $4p_{z}$ orbital to 
  (a) CaFeAsH and (b) CaFeAsF bands near E$_F$. The dashed red lined regions emphasize 
  areas of differences in bands
along the $\Gamma$-X and $\Gamma$-Z directions, see text.}
	\label{Fig3}
\end{figure}
%

Figure \ref{Fig3} shows the bands along the most relevant symmetry lines, weighted this
time by the amount of As 4p$_{z}$ character. Most prominently, in CaFeAsH, a band is
essentially linear from $\Gamma$ to Z --- very large $k_z$ dispersion -- over an energy span
of 1.5 eV. The velocity is $\hbar v_{F}$=3.8 eV$\cdot$\AA, or $v_{F}$=$5.9\times10^{8}$cm/s.
The bonding path that enables this large interlayer hopping is Fe-As-H-As-Fe, identified
previously.\cite{Muraba2014,Wang2015} 
Visible in Fig. \ref{Fig3}(b) is that the corresponding band 
in CaFeAsF has a factor of $\sim$7 less
dispersion, and in the opposite direction. Also, it lies entirely above E$_F$ so it does not
affect the FS.

A plot through the zone Z-$\Gamma$-Z would indicate this dispersive band in CaFeAsH to be
Dirac-like at -0.2 eV, except that it (necessarily, by symmetry) 
becomes quadratic extremely close to $\Gamma$.  
However, this band disperses downward in both directions in plane, so the point at $\Gamma$
is a van Hove singularity (vHs) with two normal negative masses $m_x = m_y$
and an  extremely small positive mass
along the $k_z$ direction, whereupon the band quickly becomes linear (massless). 
Lying 0.2 eV below E$_F$ this vHs has no influence on low energy
properties (including, we expect, superconductivity) but it reflects a highly unusual influence of H in
this structure. A very similar linear valence band has been found in the CoAs$_3$
class of skutterudites where it has been analyzed in detail\cite{Smith2011,Smith2012} 
and found to arise from a cluster orbital of As $p$ states surrounding the open hole
in this lattice.

There is an additional H versus F difference evident in Fig. \ref{Fig3}(b). Of the pair of hole
bands extending from $\Gamma$, the Fermi surfaces are circular and nearly identical in CaFeAsH,
while the values of $k_F$ in CaFeAsF differ. The cylinders include different numbers of
holes, and the larger number of holes in CaFeAsH are balanced by the electrons in the
ellipsoid that does not exist in CaFeAsF. This difference also affects the nesting 
function which can be seen in plots from a careful study that we present below.

\begin{figure}
	\centering
	\includegraphics[width=0.8\columnwidth]{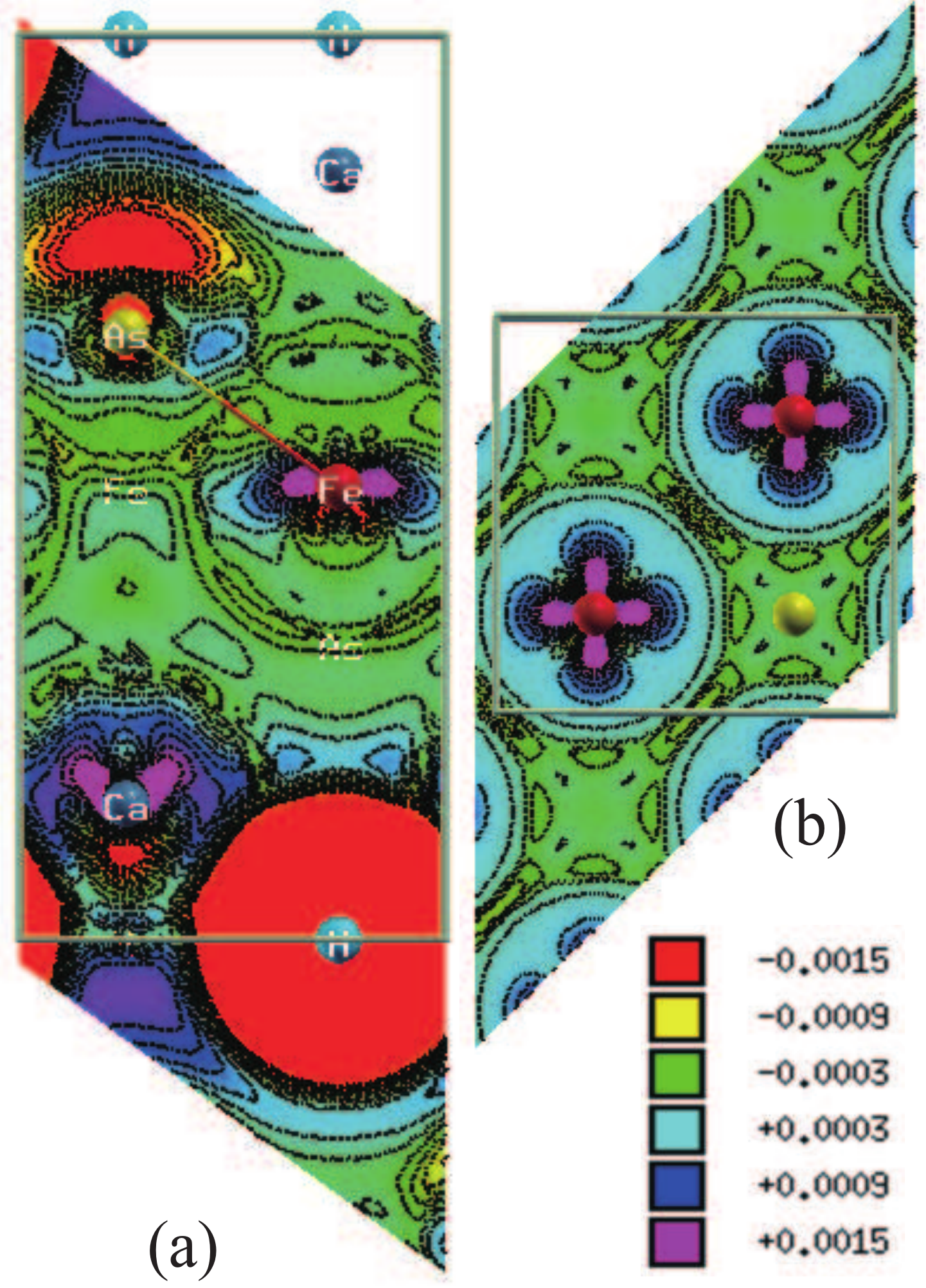}\\
	\caption{(Color online) Contour plots of the density difference
		$\rho$(CaFeAsH)-$\rho$(CaFeAsF). (a) a (110) plane containing the 
		Fe-As zigzag chain (and also the Ca and H/F sites). (b) (001) plane
		through Fe sites. The large red region in (a) denotes where contours
		have been cut off due to the large and meaningless difference of the
		H and F densities.}
	\label{Fig4}
\end{figure}

\subsection{Fe layer charge: H versus F}
Due to the focus of attention in Fe-based pnictides and chalcogenides
on Fe orbital occupations and how they relate to properties, we have
calculated the difference density $\rho$(CaFeAsH)-$\rho$(CaFeAsF),
with both calculated at the lattice constant of the former compound.
Fig. \ref{Fig4} shows  contour plots of this difference density 
in two planes: a (110) plane containing a FeAs zigzag chain, and a (001)
plane through the Fe sites. Positive values designate
regions where H attracts charge relative to F.  

In the horizontal plane containing Fe sites, Fig.~\ref{Fig4}(b), 
H is seen to increase charge in the in-plane orbitals while there is a slight decrease
in density in the interstitial region in both planes. 
In the Fe-As chain direction, Fig. \ref{Fig4}(a), the increase
in occupation of the Fe in-plane orbitals is again evident.  What is more
evident is the larger change on As, with charge ``transferred'' to the $p_x,p_y$
orbitals at the expense of the $p_z$ orbital. In addition, there are changes
on the Ca ion of the same size as on Fe. Density is
slightly decreased in the interstitial region.

These differences can be quantified by subtracting the orbital occupations,
CaFeAsH minus CaFeAsF, within the atomic spheres. In units of 10$^{-3}$,
the differences for Fe are: $d_{z^2}$, 2.5; $d_{xy}$, 7.6; $d_{x^2-y^2}$, 3.5;
$d_{xz}=d_{yz}$, -0.6. The net change for Fe is +12.4$\times$10$^{-3}$ electrons.
For As,  the analogous changes are: $p_x,p_y$, 5.9; $p_z$, -9.9, for a small net
change of +1.9$\times$10$^{-3}$ electrons. 
Thus H induces $\sim$0.015 electrons into the Fe+As spheres
compared to F, with the decrease occurring in the interstitial region or
in the Ca(H,F) region.

\subsection{Study of  H$_{1-x}$F$_x$ substitution}
\begin{figure}
	\centering
	\includegraphics[width=0.98\columnwidth]{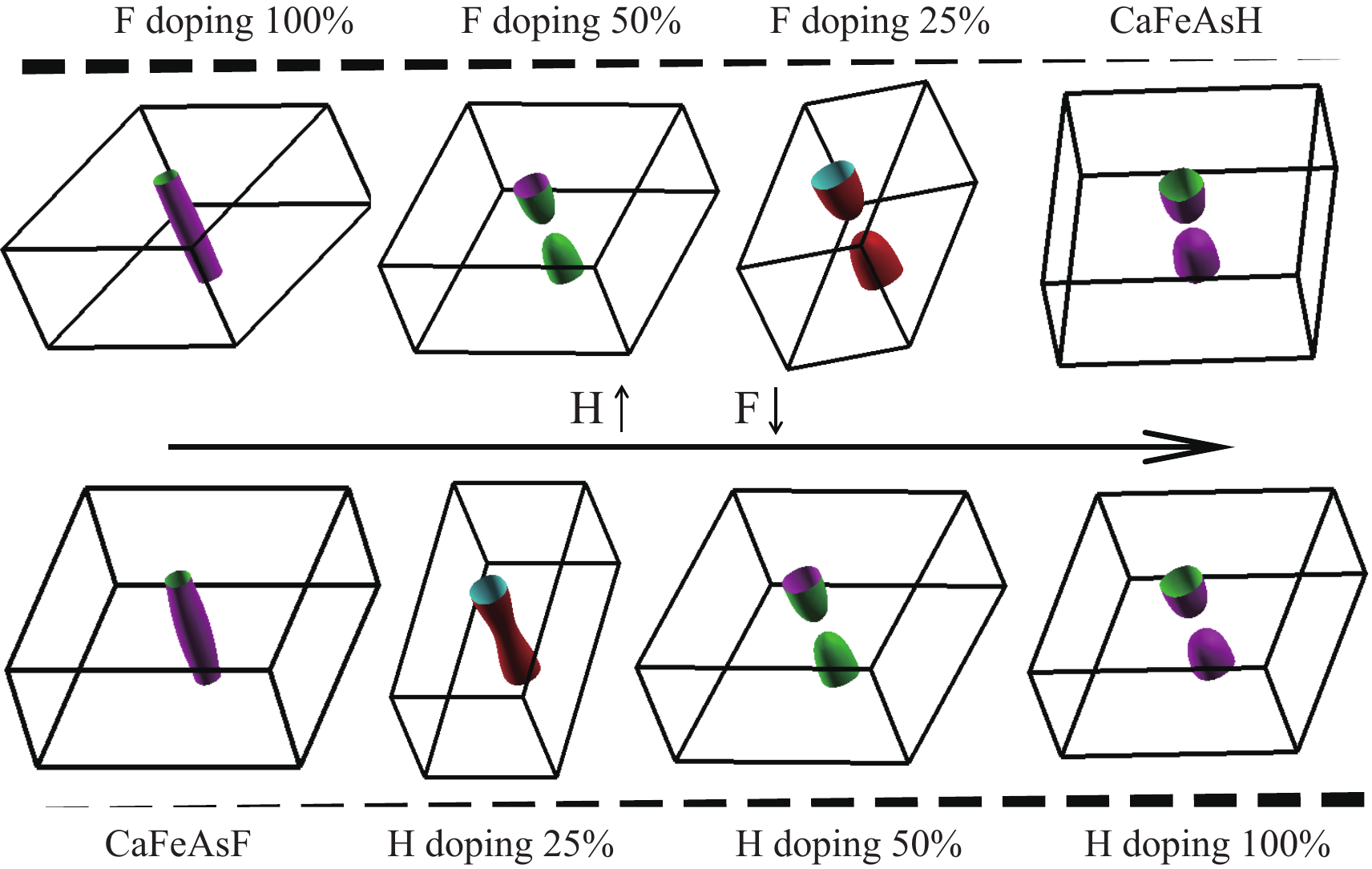}\\
	\caption{(Color online) Fermi surfaces for the indicates relative fractions of H and F.
          Upper row: F doping at the CaFeAsH structural parameters. Lower row: H doping 
          at the CaFeAsF structural parameters.}
	\label{Fig5}
\end{figure}

To determine whether the Fermi surface differences (and band structure differences)
are affected by the  different lattice parameters 
we have replaced H by F in CaFeAsH and replaced F by H in CaFeAsF, each with its
own value of the $c$ lattice parameter.\cite{Hosono2014, Matsuishi2009} The differences 
in the 3D ellipsoid FS sheet are shown in Fig.~\ref{Fig5}, 
replacing H by H$_{1-x}$F$_{x}$ in CaFeAsH, for $x$=1, $\frac{1}{2}$, 
$\frac{1}{4}$, 0,  and replacing F by H$_{y}$F$_{1-y}$ in CaFeAsF 
($y$=0, $\frac{1}{4}$, $\frac{1}{2}$, 1).
The effect of $c$ lattice constant is very small; note that in each of the cases
$x = 0, \frac{1}{2},$ and $1$ the surfaces are very similar, with the pinching
off of the cylinder occurring around 35-40\% H. Evidently the difference is a
chemical bonding one that is much more important than intersublayer separations.

%



%
\begin{figure}
  \centering
  \includegraphics[width=0.98\columnwidth]{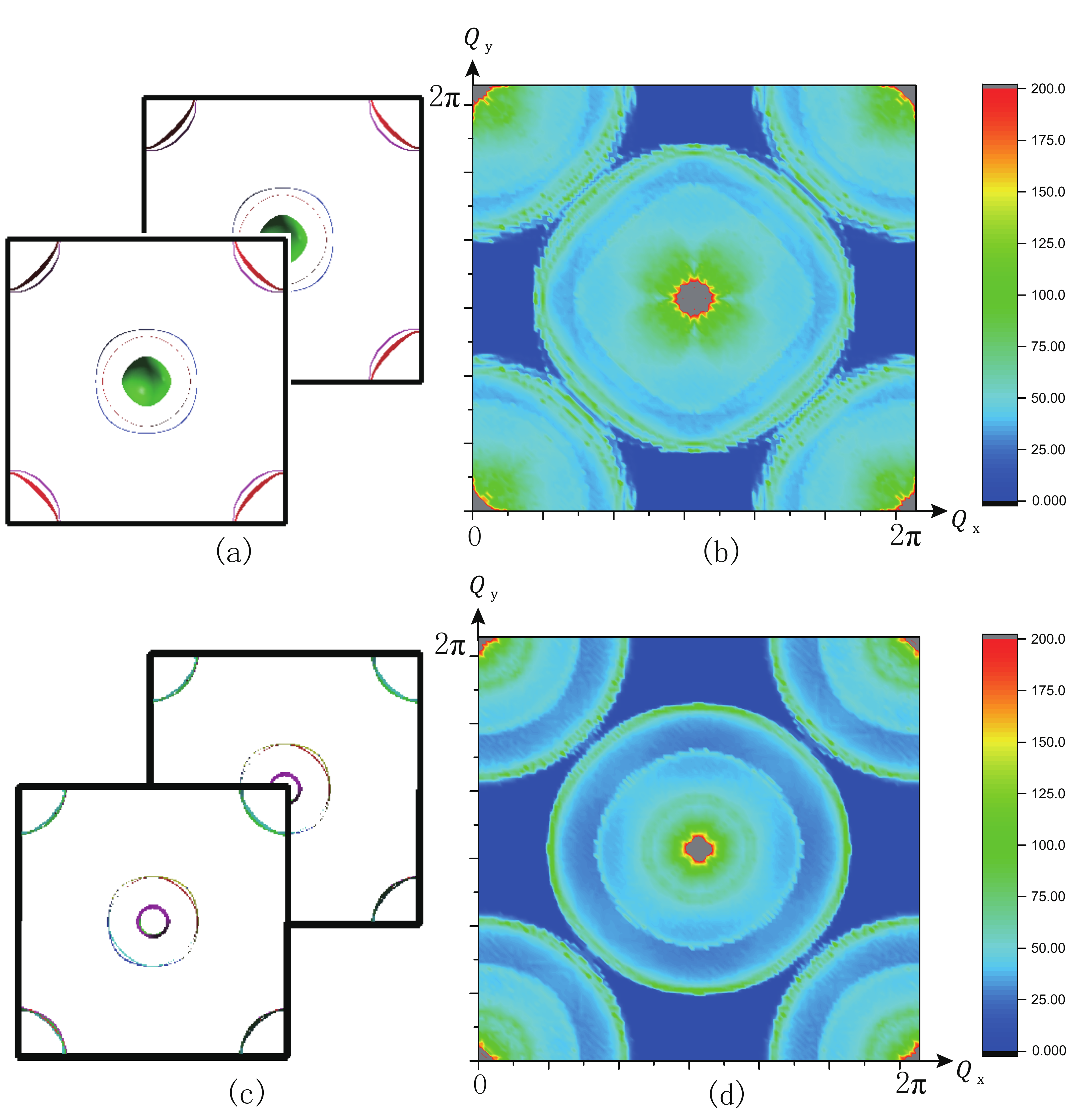}\\
  \caption{(Color online) Left column: top view of Fermi surface sheets of nonmagnetic
  (a) CaFeAsH and (c) CaFeAsF. The FS contours are overlaid by a copy displaced by 
$\mathbf{Q} =(\pi,\pi,0)$. The close overlap of hole and electron pockets indicates 
strong Fermi surface nesting. Right Column: the $\mathbf{q}$-space plot of 
  the nesting function $\xi(\mathbf{q})$ 
of (b) CaFeAsH and (d) CaFeAsF as described in the text. The plot covers the entire 
zone, with the plot centered at $\mathbf{Q}=(\pi,\pi,0)$; $\Gamma$ points lie at the corners. 
The structures are described in the text.}
\label{Fig6}
\end{figure}

\subsection{Fermi surface nesting}
Fermi surface nesting has played a central role in theories of the Fe-based
superconductors, on the one hand accounting for the observed AFM phases in
several undoped compounds and on the other as promoting short-range AFM
fluctuations that may provide the superconducting pairing mechanism. Both
of these aspects focus on the susceptibility peak near $\vec Q = (\pi,\pi,0)$,
as discussed for Fe pnictides earlier.\cite{Zhang2010,Liu2012}
However, nesting processes, {\it i.e.} those which scatter from Fermi 
surface to Fermi surface,
dominate all low energy processes, and it may be important to understand such
process throughout the Brillouin zone.
  
With an eye first toward the magnetic instability in CaFeAsH and 
CaFeAsF and the wavevector dependence of magnetic fluctuations in the 
(high temperature) tetragonal structure, 
we analyze the Fermi
surface topology in the nonmagnetic phase. The top view of Fermi surface sheets is shown
in Fig.~\ref{Fig6}(a) and Fig.~\ref{Fig6}(c), with a copy displaced by 
$\vec Q$. The cylindrical,
nearly circular, FS sheets at M, and two of the three at $\Gamma$,
reflect the  2D character of these Fermi surfaces. As mentioned, a substantial
degree of $\vec Q$ nesting is evident, which we proceed to quantify below.
While the circular FSs around $\Gamma$ in CaFeAsH essentially coincide, in the
F compound the inner sheet is much smaller, arising from the difference
in band structures discussed above.

\subsubsection{Formalism}
A quantitative measure of nesting is provided by the
Fermi surface nesting function $\xi(\vec q)$, which measures the phase space for
scattering through wavevector $\vec q$ from $\vec k$ on the FS to $\vec k + \vec q$
also on the FS. This function is given by

\begin{eqnarray}
 \xi_{\bf{q}} &=& \sum_k\delta(\varepsilon_{\vec k}-\varepsilon_F)
                       \delta(\varepsilon_{\vec k+\vec q}-\varepsilon_F) \nonumber\\
&=& \frac{\Omega_c}{(2\pi)^3}\int_{\mathcal{l}} \frac{d\mathcal{L}_{\vec k}(\varepsilon_F)}
          {|\vec v_k \times \vec v_{k+q}|}
\label{nestingeq}
\end{eqnarray}
here $\Omega_c$ is the cell volume and $\vec v_k = \nabla_k \varepsilon_k$ is
the electron velocity ($\hbar$=1).
The latter expression, after the volume integral with two $\delta$-function
restrictions,  gives a geometrical interpretation: it is the integral over the line
of intersection $\mathcal{L}_{\bf{k}}(\varepsilon_F)$
of the undisplaced FS and a copy of the FS displaced by $\vec q$, weighted by
the inverse of the cross product of the two velocities. Large contributions arise
from regions of parallel or antiparallel velocities, {\it i.e.}  FS nesting where the velocity
cross product becomes small.
Contributions are enhanced by small
velocities, although that is not an issue with these cylindrical FSs that arise
in these compounds, with near constant Fermi velocities around the FSs. $\xi(\vec q)$ has a trivial 
$q^{-1}$ divergence at $q\rightarrow$0 that
is countered by vanishing matrix elements in physical properties involving
such long wavelength momentum transfers.

Fermi surface instabilities are more often analyzed in terms of the electronic
susceptibility, which is the relevant response function,
\begin{align}
 {{\chi }_{0}}\left( \mathbf{q} \right)=\frac{1}{N}\sum\limits_{k,m,n}{|M_{kn,k+q~m}|^2}{
 \frac{f\left( {{\varepsilon }_{\vec k + \vec q,m}}
 \right)-f\left( {{\varepsilon }_{\vec k,n}}
 \right)}{{{\varepsilon }_{\vec k + \vec q,m}}-
 {{\varepsilon }_{\vec k,n}}+i\eta}}.
\end{align}
where $M_{km,k^{\prime}n}$ is a matrix element of $exp(i\vec q \cdot \vec r)$ between
Bloch functions. Though matrix elements often are important, evaluations without
matrix elements, corresponding to the ``generalized susceptibility,'' are
much more common. Peaks in the real part of $\chi_0(\vec q)$ provide the
positions of potential wavevector $\vec q$ instabilities. Structure in $\chi_0(\vec q)$
arises from Fermi surface nesting as well as from other near-E$_F$
structure in the band structure, viz. the virtual bands that will be discussed below. 
  
Instabilities in the metallic phase, often phrased as instabilities of the Fermi surface, 
are characterized in terms of peaks in the real part of the
electronic susceptibility. 
The nesting function defined above has its own separate importance. 
To first (linear) order in small $\omega$ the connection to the
imaginary part $\chi_0''$ of the dynamical susceptibility 
(here matrix elements are set to unity) is
\begin{equation}
 \chi''_0(\vec q,\omega) = \pi\omega \xi(\vec q) + {\cal O}(\omega^2).
\end{equation}
Thus low energy dissipative processes occur where there is strong weight in
$\xi(\vec q)$,
in addition to the Drude interband absorption at small $|\vec q|$.

\subsubsection{Discussion of nesting} 
The behavior of $\xi(\vec q)$ in both compounds is dominated by the very strong 
peak precisely at $\vec Q$.
In several Fe-based pnictides a maximum at this wave vector also shows up in
$\chi(\vec q,\omega=0)$ but probably not with nesting as perfect as in this
F compound, where Fig.~\ref{Fig6}(c) illustrated the perfect nesting of the
outer electron and hole surfaces in CaFeAsF. The nesting is not as perfect for CaFeAsH,
but there are two surfaces of each type of carrier that nest. Note that in the dark
regions of these plots there is zero nesting, there are no low energy
excitations in these regions, a result of the relatively small Fermi surfaces.   
Of interest, of course, is that the near perfect nesting at 
$\vec Q$ indicates a strong tendency for AFM order,
and that (electron) doping will degrade the nesting and thereby
diminish the tendency for magnetic ordering.

Additional structure can be seen in 
$\xi(\vec q)$ that is difficult to see in the susceptibility. In 2D, a circular FS of 
radius $k_F$ gives rise to a characteristic divergence\cite{WEP1,WEP2} 
$1/\sqrt{2k_F-q}$ as $q = |\vec q|$ approaches
$2k_F$, then vanishes abruptly for $q>2k_F$ where transitions are no longer available. The essentially
identical in size circular FSs of CaFeAsF give rise to this characteristic circular peak at radius of 
$2k_F$ around $\vec q$=0 and almost identically around $\vec q = \vec Q$, 
easily visible in the figure. Since phonons acquire renormalization from small (nearly zero) 
energy processes, $\xi(\vec q)$ suggests strong phonon renormalization at the nesting 
wavevector and smaller but definite Kohn anomalies
on the $q\approx 2k_F$ and $|\vec q -\vec Q|\approx 2k_F$ circular
ridges. Spinwaves should acquire 
renormalization in the same regions, for the same reasons.  

Figure~\ref{Fig6}(b) provides the nesting function for CaFeAsH in comparison to that of
CaFeAsF in Fig.~\ref{Fig6}(d). In the H compound, the non-circular nature of the ridges of 
$\xi(\vec q)$ reflect the not-quite-circular Fermi surfaces. In addition, there is
additional structure near the peaks at $\Gamma$ and $\vec Q$ that arise from the ellipsoidal
Fermi surface. Taking the broader picture, however, there is little quantitative difference between
the nesting functions of these isoelectronic compounds.

\subsection{Antiferromagnetic phases}

Here we discuss magnetic energies and the basic features of the electronic structure of
the SAFM phase in undoped CaFeAsH, and draw some parallels with related pnictides.
The types of AFM order that are studied are described in Sec. II.B, some
of which require a $\sqrt{2}$$\times$$\sqrt{2}$$\times$1 supercell (there are two Fe sites
within the primitive cell).
The energies of four of the simplest magnetic configurations relative to the nonmagnetic phase are
presented in Table I, where the Fe atomic sphere moments are also provided.
It has been much discussed that Fe moments are overestimated by DFT 
methods,\cite{Ortenzi,Han,Seb} so
magnetic energies will also be exaggerated, however relative orderings may still be
meaningful.  These relative energies are in almost the same order as in the related
compound CaFeAs$_2$, but differ somewhat in magnitude,\cite{Huang2015,Liu2011} reflecting
some sensitivity to the Fe-As structure or to the makeup of the blocking layer. 

\begin{table*}
\centering
\caption{Total energy difference, for CaFeAsH, of five magnetic phases including
nonmagnetic (NM), ferromagnetic (FM), N{\'e}el AFM (NAFM), striped
AFM (SAFM) and bi-striped AFM (BSAFM). The reference is the energy of
the NM phase ($\Delta$$E$=$E$
$_{(A)FM}$-$E$$_{NM}$). The corresponding magnetic moment in the Fe sphere is given.}
\begin{tabular}{c|>{\centering}p{1.5cm}>{\centering}p{1.5cm}>{\centering}p{1.5cm
}>{\centering}p{1.5cm}>{\centering}p{1.5cm}>{\centering}p{1.5cm}}
\hline
\hline
  Magnetic structure    &    NM   &    FM   &   NAFM   &  SAFM  &
  BSAFM \tabularnewline
\hline
 (a) Relative energy (meV) & $0$ & $-83$ & $-119$ & $-282$ &
 $-163$  \tabularnewline
\hline
\hline
 (b) Fe moment (${{\mu }_{B}}$) & $0$ & $0.63$ &$1.92$ & $1.92$ & $1.22$ \tabularnewline
\hline
\end{tabular}
\end{table*}

\begin{figure}
  \centering
  \includegraphics[width=0.98\columnwidth]{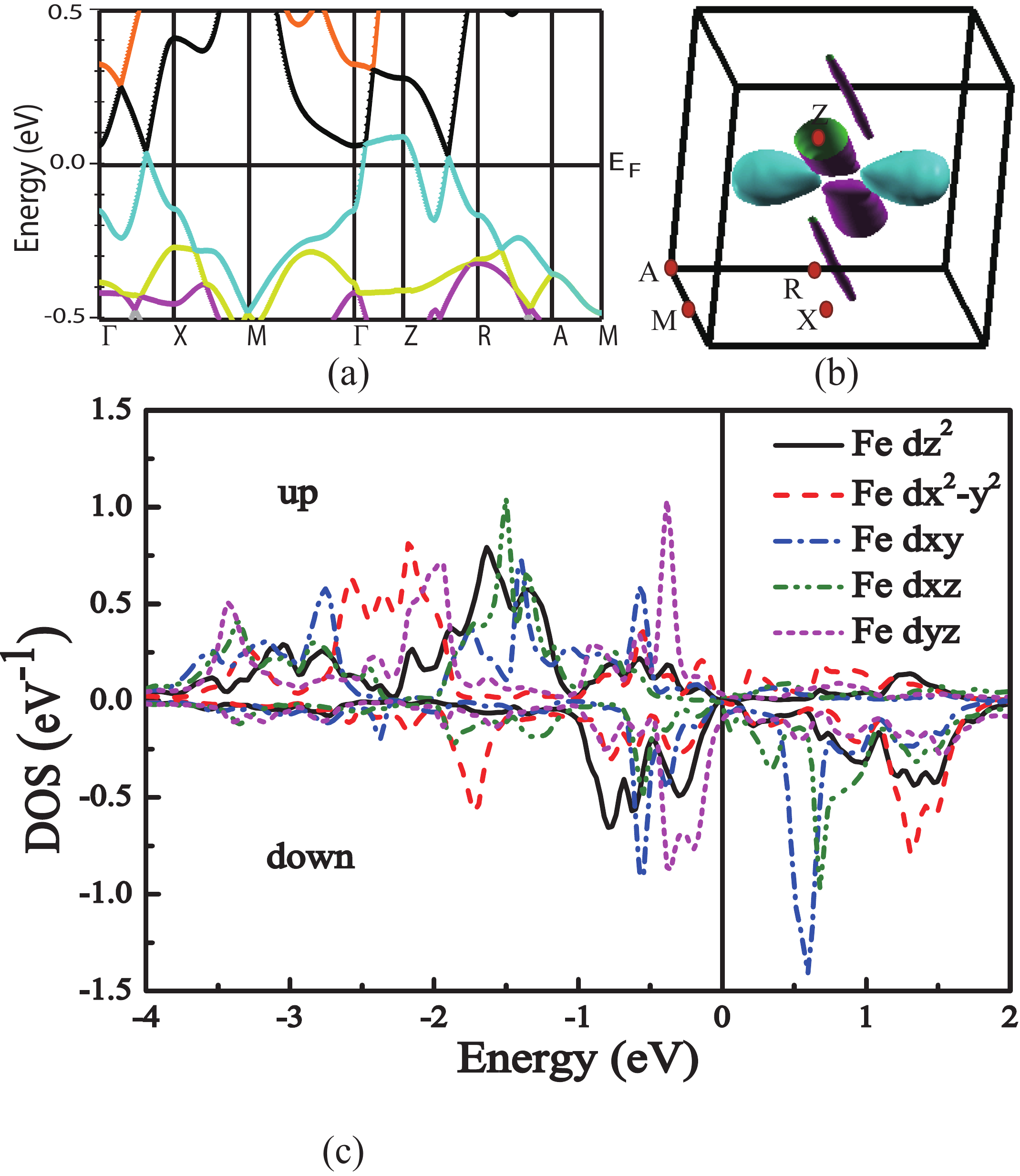}\\
  \caption{(Color online) For CaFeAsH with SAFM order: (a) electronic band structure,
(b) density of states, and (c) projected DOS for the Fe atom. Note the small
DOS at the Fermi energy (zero of energy).}
\label{Fig7}
\end{figure}

For FM alignment
the moment is much smaller and the energy less favorable than for the AFM alignments. The smaller
moments account qualitatively for the FM phase being less favorable.
The SAFM state is most favorable by a large margin, 120 meV/Fe lower than the
BSAFM phase. This lowest energy holds in spite of the SAFM phase having the same Fe spin moment of
1.92 $\mu_{B}$ as does the NAFM case. The differences in Fe moments in the various phases is evidence
that the magnetism has substantial itinerant character. For itinerant magnets the energy gain
for magnetism is $Im^2/4$, where $I$ is the Hund's exchange coupling (also called Stoner $I$)
and $m$ is the moment. This
behavior is not observed, arising possibly from non-linear effects
but also that the moment may not be of the simplest itinerant type.

For comparison, the SAFM moment of Fe is 2.18 $\mu_{B}$ in CaFeAsF, 13\% higher than in
CaFeAsH. Thus the differences in electronic structure discussed above significantly affect the magnetic
predictions in these compounds. In CaFeAsF, the reported experimental 
moment is 0.49 $\mu_{B}$,\cite{Mishra2011} less than 25\% of the calculated value. This discrepancy follows
the known substantial overestimate of
the moment in these Fe-based materials by DFT calculations. 
To our knowledge, no experimental moment has been 
reported for CaFeAsH, so it is unclear whether the differences in band structure between the H and
F compounds shows up experimentally.

Figure~\ref{Fig7} shows the CaFeAsH SAFM band structure, Fermi surfaces, and projected
Fe DOS. The magnetic ordering results in an orthorhombic cell, with a very low Fermi
level density of states.  The bands near E$_F$ at $\Gamma$
and M have been mixed and gapped by the SAFM order, 
so the cylindrical FSs have vanished,
reflecting the massive electronic reconstruction driven by the nesting.  
The few remaining bands crossing E$_F$ have high velocity.

The resulting electronic structure has noteworthy aspects. The majority states
are filled ($d^5)$. The minority states seem at first sight to have an
expected $e_g-t_{2g}$ splitting, however it is more complex than that.
Evidently the SAFM order affects substantially the Fe configuration, based on
the loss of tetrahedral symmetry due to magnetic reconstruction
of the electronic bonding. There are substantial peaks of minority $d_{x^2-y^2}$
states both at -2 eV and +1.5 eV. The main $d_{xy}$ peak is above E$_F$ but a
noticeable fraction lies below E$_F$. $d_{xz}$ is mostly above E$_F$; $d_{yz}$ is
entirely below E$_F$. The magnetic order thus has a considerable effect on the
orbital occupations and thereby the electronic structure.

\subsection{La and Co Doping Phase}
\begin{figure}
  \centering
  \includegraphics[width=0.9\columnwidth]{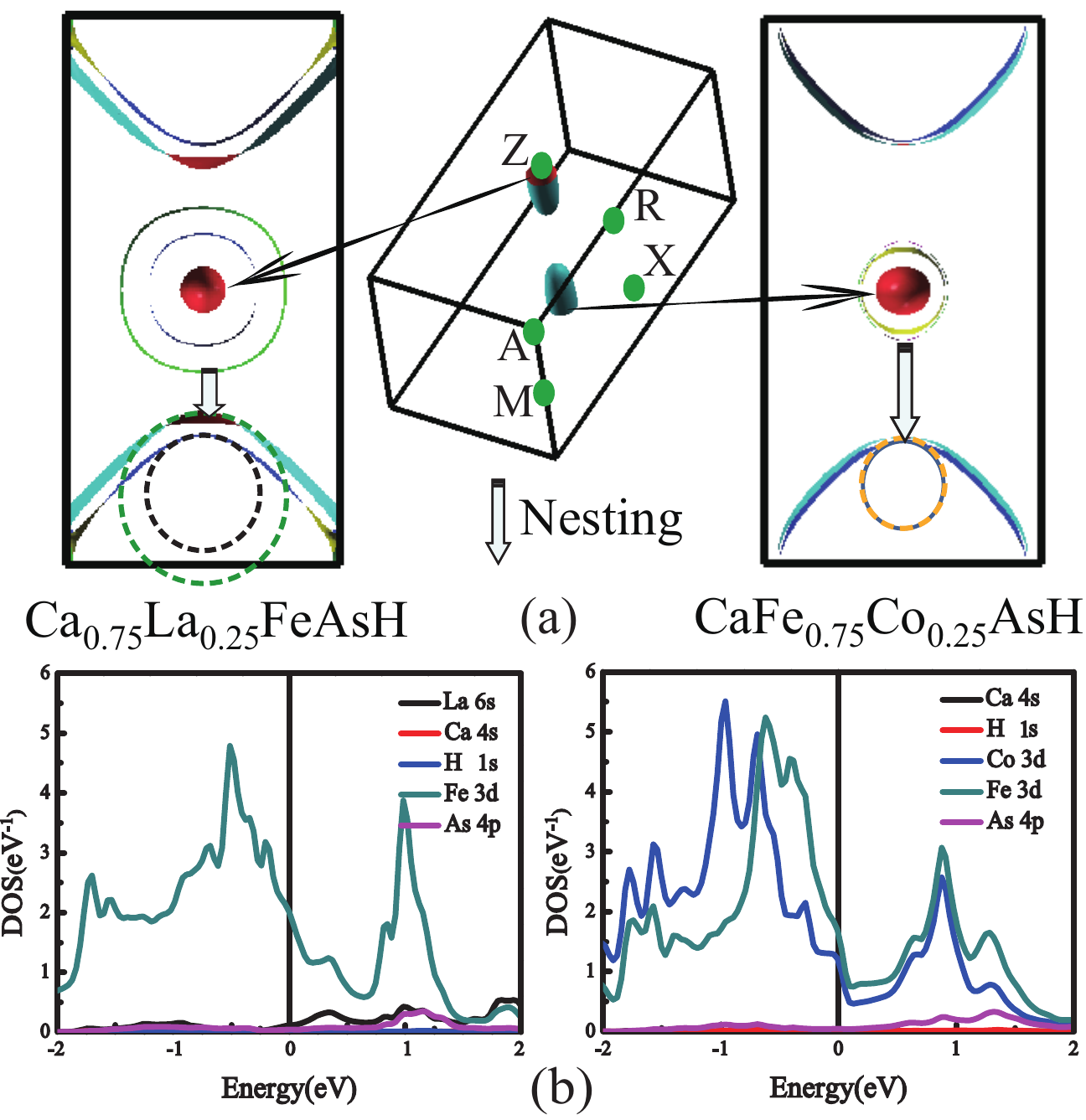}\\
  \caption{(Color online) (a) The Fermi surfaces and qualitative nesting features of 
   Ca$_{0.75}$La$_{0.25}$FeAsH and CaFe$_{0.75}$Co$_{0.25}$AsH, with
   electron doping in each case but indirect versus direct respectively. 
   (b) Density of states (DOS) of each atom in Ca$_{0.75}$La$_{0.25}$FeAsH 
  (where only Fe is significant)
   and CaFe$_{0.75}$Co$_{0.25}$AsH, where Co can be compared with Fe.}
  \label{Fig8}
\end{figure}

Both La and Co provide electronic doping, however Co introduces changes
directly into the Fe layer while La simply donates electrons.
Here we analyze the differences in electronic structure between indirect (La doping) and
direct (Co doping),\cite{Muraba2014, Hosono2014} to identify aspects that may correlate
with the question of why T$_{c}^{max}$ is a factor of two different in the two cases. 
We present results for the same 25\% electron doping of the DOS and different Fermi 
surfaces of Ca$_{0.75}$La$_{0.25}$FeAsH and 
CaFe$_{0.75}$Co$_{0.25}$AsH in Fig.~\ref{Fig8}(a).
The  electron surfaces of course grow in size while hole surfaces decrease, and with
substantial changes. We note that the onset T$_c$ in the La-doped
case remains around 47 K from $x$=0.08 (metallization) to $x$=0.30, the
highest doping level reported. This makes our 25\% case representative.
The data indicate an unusual insensitivity to doping level,
unlike the more common superconducting dome behavior with doping.
It may however be possible that there is mesoscopic chemical phase separation in these
sample such that the onset T$_c$ represents the value of a single phase.
For the homogeneous phases that we are modeling, there are substantial changes with doping.

The projected DOS curves of Fig.~\ref{Fig8}(b) reflect large
differences. For La doping the Fe DOS hardly changes, with E$_F$ increasing
rigid-band like. The band filling of course is the same for Co doping.
The Co $3d$ DOS differs from that of Fe, however, with the main peak
lying 0.5 eV lower (higher binding energy), with the result that the
contribution to N(E$_F$) is substantially smaller. For this case E$_F$
lies right at the top of a band, where N(E) drops discontinuously. The
H contribution at E$_F$ is very small in both cases. The value of N(E$_F$)
is similar in the two cases, but disorder broadening may lower the value
in the Co-doped case due to the step edge in N(E). 

The 3D ellipsoid pocket remains in each of these cases. Because the velocity
along $k_z$ is large, the extent of the ellipsoid along $k_z$ does not
increase much, being larger around the waist and larger  
in the Co doping case. 
The nesting of electron and hole FSs is severely degraded by 25\% doping,
a change that is conventionally associated with the
destruction of spin density wave order that enables superconductivity to emerge.
Although Cooper pairing due to short-range AFM fluctuations that are
encouraged by nesting provides the most common explanation for superconductivity
in Fe pnictides, there is no longer great interest in details of degraded
nesting and we do not provide $\xi(\vec q)$ for the doped materials.
Fig.~\ref{Fig8}(a) provides some indication of where some nesting remains,
by repeating the electron FS cylinders displaced by the most evident
displacement. The nesting lies at a larger displacement for Co-doping
and displays weaker nesting (the corresponding curvatures of the FSs differ
more). Such differences might contribute to the factor of two difference
in T$_c$, but the theory is not sufficiently established to pursue
this point.

\section{SUMMARY}

Density functional based methods have been applied to assess various aspects of the
effect of Ca and H, versus La and O, on the underlying electronic structures
of these 1111 compounds. Secondarily, the differences in electron doping,
by La for Ca or by Co for Fe, have been modeled and analyzed.
The isovalent system CaFeAs(H,F) has also been studied.
The key feature introduced by H is an unusual band dispersing along $k_z$, as
pointed out previously, due to a Fe-As($p_z$)-H-As($p_z$)-Fe hopping path enabled
by the hydrogen $1s$ orbital that is much less strongly bound than the corresponding
$2p$ orbitals in the F compound. 

This peculiar band introduces an additional Fermi surface of three dimension
character and ellipsoidal shape, something not appearing in other 1111 compounds
including the isovalent F compound. Both compounds show very strong Fermi surface
nesting at the expected $(\pi,\pi,0)$ wavevector. The cylindrical Fermi surfaces
are different for the two compounds: circular for the F compound but two surfaces
with different radii; not so circular for the H compound but nearly identical
surfaces. The unusual band and extra Fermi surface has been discussed, briefly,
in terms of the effect of an incipient band that could help to resolve the
differences observed between these two compounds. The differences in orbital
occupations in both the Fe and As atomic spheres in the H and F compounds has
been quantified. It may be helpful in future to compare these isovalent substations
with the results on H-doping of the 1111 class, as in 
LaFeAsO$_{1-x}$H$_x$\cite{Hosono2013} and 
SmFeAsO$_{1-x}$H$_x$.\cite{Takahashi2012}

The effects of indirect electronic doping (La for Ca) and direct doping (Co for Fe)
have been contrasted. The effect on the density and character of 
states near the Fermi level is
substantial, so differences in properties, including T$_c$, are expected. 
Differences in the nesting of these two
electron-doped phases have been described as well.
However, with no material-specific theory of T$_c$ in Fe-based superconductors,
it would be speculation to try to identify the most important distinctions. 

We expect that these several findings will contribute to investigations into
the determination of what aspects of these differences are relevant to understand
their superconducting distinctions. The observation of a lack of superconducting
dome in the phase diagrams of both types of doping, taken at face value, indicates
an indifference of superconductivity to details of the electronic structure that
is difficult to comprehend, since many studies in Fe-based superconductors
point to important effects due to small differences.   

\section{ACKNOWLEDGMENTS}

We acknowledge useful discussions with X. H. Chen on their experimental data, and
helpful comments from X. L. Yu, W. C. Bao, A. S. Botana,
and S. Gangopadhyay. This work was 
supported by NSFC of China under Grant Nos. 11274310, 11474287 and 11574315. 
The calculations were performed at the Center for Computational Science of CASHIPS, 
the ScGrid of Supercomputing Center and Computer Network Information Center of 
Chinese Academy of Science.
W. E. Pickett was supported by US National Science Foundation award DMR-1207622.

\end{document}